\documentclass[prd,showpacs,preprintnumbers,amsmath,amssymb]{revtex4}

\usepackage{graphicx}
\usepackage{dcolumn}
\usepackage{bm}

\def\be{\begin{eqnarray}}
\def\ee{\end{eqnarray}}
\def\bea{\begin{eqnarray}}
\def\eea{\end{eqnarray}}

\def\pp{{\bf p}_\perp}

\def\xT{{\bf x}_\perp}

\def\bT{{\bf b}_\perp}
\def\bp{{\bf b}_\perp}

\def\RT{{\bf R}_\perp}

\def\0T{{\bf 0}_\perp}
\begin{document}


\title{Parton Orbital Angular Momentum and Final State Interactions}

\author{Matthias Burkardt}
 \affiliation{Department of Physics, New Mexico State University,
Las Cruces, NM 88003-0001, U.S.A.}
\date{\today}

\begin{abstract}
Definitions of orbital angular momentum based on Wigner distributions
are used as a framework to discuss the connection between the Ji definition
of the quark orbital angular momentum and that of Jaffe and Manohar.
We find that the difference between these two definitions can be interpreted as the change in the quark orbital angular momentum as it leaves the target in a DIS experiment.
The mechanism responsible for that change is similar to the mechanism that
causes transverse single-spin asymmetries in semi-inclusive deep-inelastic scattering.
\end{abstract}

\maketitle
\section{Introduction}
Generalized Parton Distributions (GPDs) have been identified as a powerful tool to analyze the
angular momentum decomposition of the nucleon \cite{JiPRL}.
Furthermore GPDs can also be used to create truly three-dimensional images of the nucleon in the form of impact parameter dependent parton distributions \cite{mbGPD}. These images in a space where one dimension describes the light-cone
momentum fraction and the other two dimensions describe the transverse position of the parton (relative to the
transverse center of momentum) are complemented by Transverse Momentum dependent
parton Distributions (TMDs) \cite{ams}. Wigner distributions provide a framework that allows a simultaneous description of GPDs and TMDs \cite{wigner}.

Orbital Angular Momentum (OAM) correlates the position and momentum of partons. One can thus utilize Wigner distributions, which
simultaneously embody the distribution of position and momentum,  to define OAM \cite{lorce,jifeng}.  
However, in the definition of these distributions,
care must be applied to ensure manifest gauge invariance. In general, this can be accomplished by connecting any nonlocal correlation function with a Wilson-line gauge link. Specifying a
Wilson-line gauge link requires selecting a path along which the vector potential is evaluated.
The choice of path raises the immediate issue of how the quantities defined using
Wigner distributions (TMDs, OAM, ...) depend on that choice. The importance of this issue had become evident in the context of Single-Spin Asymmetries (SSAs) \cite{BHS}. Indeed, while a straight-line gauge link definition of TMDs yields a vanishing Sivers effect \cite{sivers,collins1}, the correct gauge link relevant for TMDs in
Semi-Inclusive Deep-Inelastic Scattering (SIDIS) involves a detour to light-cone infinity
\cite{jifengTMD} in order to properly include final-state interactions. In light-cone gauge, this subtlety had first been overlooked since in that gauge the Sivers effect solely arises from the contribution from the gauge-link piece at light-cone infinity \cite{jifengTMD}.

With Wigner distributions and OAM defined through them these issues arise all over again
\cite{jifeng,hatta,lorce2}. The main goal of this note is to address that dependence of OAM defined through Wigner distributions on the choice of path for the gauge link and 
to interpret the resulting difference between common definitions
of OAM.

\section{Angular Momentum Decompositions}

Since the famous EMC experiments revealed that only a small fraction
of the nucleon spin is due to quark spins \cite{EMC}, 
there has been a great
interest in `solving the spin puzzle', i.e. in decomposing the
nucleon spin into contributions from quark/gluon spin and
orbital degrees of freedom.
In this effort, the Ji decomposition \cite{JiPRL}
\begin{equation}
\frac{1}{2}=\frac{1}{2}\sum_q\Delta q + \sum_q { L}_q^z+
J_g^z
\label{eq:JJi}
\end{equation}
appears to be very useful: through GPDs,
not only the quark spin contributions $\Delta q$ but also
the quark total angular momenta $J_q \equiv \frac{1}{2}\Delta q + 
{ L}_q^z$ (and by subtracting the spin piece also the
the quark orbital angular momenta $L_q^z$) entering this decomposition
can be accessed experimentally.
The terms in (\ref{eq:JJi}) are defined as expectation
values of the corresponding terms in the angular momentum tensor
\begin{equation}
M^{0xy}= \sum_q \frac{1}{2}q^\dagger \Sigma^zq +
\sum_q q^\dagger \left({\vec r} \times \frac{1}{i}{\vec D}
\right)^zq
+  
\left[{\vec r} \times \left({\vec E} \times {\vec B}\right)\right]^z
\label{M012}
\end{equation}
in a nucleon state polarized in the $+\hat{z}$ direction. Here
${\vec D}={\vec \partial}-ig{\vec A}$ is the gauge-covariant
derivative.
The main advantages of this decomposition are that each term can be 
expressed as the
expectation value of a manifestly gauge invariant
local operator and that the
quark total angular momentum $J^q=\frac{1}{2}\Delta q+L^q$
can be related to GPDs 
\cite{JiPRL} 
and is thus accessible in deeply virtual Compton scattering and
deeply virtual meson production and can also be
calculated in lattice gauge theory. 
Recent lattice calculations of GPDs \cite{lattice} yielded the surprising result that
the light quark orbital angular momentum (OAM) is consistent with
$L^u\approx -L^d$, i.e. $L^u+L^d\approx 0$. Unless there is
a large contribution from disconnected quark loops, 
that had been  so far omitteed, this
would imply that $J^g \approx \frac{1}{2}\cdot 0.7$ represents
the largest piece in the nucleon spin decomposition. 

Jaffe and Manohar have proposed an alternative decomposition of the
nucleon spin, which does have a partonic interpretation
\cite{JM}, and in which also two terms, $\frac{1}{2}\Delta q$ and $\Delta G$,
are experimentally accessible
\begin{equation}
\frac{1}{2}=\frac{1}{2}\sum_q\Delta q + \sum_q {\cal L}^q+
\Delta G + {\cal L}^g.
\label{eq:JJM}
\end{equation}
The individual terms in (\ref{eq:JJM}) can be defined as matrix elements of the corresponding
terms in the $+12$ component of the angular momentum tensor
\begin{equation}
M^{+12} = \frac{1}{2}\sum_q q^\dagger_+ \gamma_5 q_+ +
\sum_q q^\dagger_+\left({\vec r}\times \frac{1}{i}{\vec \partial}
\right)^z q_+  
+ \varepsilon^{+-ij}\mbox{Tr}F^{+i}A^j
+ 2 \mbox{Tr} F^{+j}\left({\vec r}\times \frac{1}{i}{\vec \partial} 
\right)^z A^j
\label{M+12}
\end{equation}
for a nucleon polarized in the $+\hat{z}$ direction.
The first and third term in (\ref{eq:JJM},\ref{M+12}) are the
`intrinsic' contributions (no factor of ${\vec r}\times $) 
to the nucleon's angular momentum $J^z=+\frac{1}{2}$ and have a 
physical interpretation as quark and gluon spin respectively, while
the second and fourth term can be identified with the quark/gluon
OAM.
Here $q_+ \equiv \frac{1}{2} \gamma^-\gamma^+ q$ is the dynamical
component of the quark field operators, and light-cone gauge
$A^+\equiv A^0+A^z=0$ is implied. 
The residual gauge invariance can be fixed by
imposing anti-periodic boundary conditions 
${\vec A}_\perp({\bf x}_\perp,\infty)=-
{\vec A}_\perp({\bf x}_\perp,-\infty)$ on the transverse components
of the vector potential.
${\cal L}$ also naturally arises in a light-cone wave function description of
hadron states, where $\frac{1}{2}=\frac{1}{2}\sum_q \Delta q + \Delta G+
{\cal L}$, in the sense of an eigenvalue equation, is manifestly satisfied for each Fock component individually \cite{LCWF}.

A variation of (\ref{eq:JJi}) has been
suggested in Ref. \cite{Wakamatsu}, where part of $L_q^z$ is attributed to the
glue as 'potential angular momentum'. As we will discuss in the following, the potential angular momentum also has a more physical interpretation as the effect
from final state interactions.
Other decompositions, in which only one
term is experimentally accessible, will not be discussed in this brief note.

\section{TMDs and Orbital Angular Momentum from Wigner Distributions}
Wigner distributions can be defined as 
defined as off forward matrix elements of non-local
correlation functions \cite{wigner, jifeng,metz}
\begin{equation}
W^{\cal U}(k^+=xP^+,{\vec b}_\perp, {\vec k}_\perp)
\equiv \int \frac{d^2{\vec q}_\perp}{(2\pi)^2}\int \frac{d^2\xi_\perp d\xi^-}{(2\pi)^3}
e^{-i{\vec q}_\perp \cdot {\vec b}_\perp}
e^{i(xP^+\xi^-\!-{\vec k}_\perp\cdot{\vec \xi}_\perp)}
\langle P^\prime S^\prime |
\bar{q}(0)\Gamma {\cal U}_{0\xi}q(\xi)|PS\rangle
\label{eq:wigner}
\end{equation}
with $P^+=P^{+\prime}$, $P_\perp = -P_\perp^\prime = \frac{q_\perp}{2}$.
Throughout this paper, we will chose ${\vec S}={\vec S}^\prime = \hat{\vec z}$. Furthermore, we will focus on the 'good' component by selecting $\Gamma=\gamma^+$.
In order to ensure manifest gauge invariance, a Wilson line gauge link 
${\cal U}_{0\xi}$ connecting the quark field operators at position $0$ and $\xi$ must 
be included. The issue of choice of path
for the Wilson line will be addressed below. 

In terms  of  Wigner distributions, quark transverse momentum and OAM can be defined respectively as \cite{lorce}
\begin{eqnarray}
\langle {\vec k}_\perp\rangle_{\cal U} &=&
\int dx d^2{\vec b}_\perp d^2{\vec k}_\perp {\vec k}_\perp 
W^{\cal U}(x,{\vec b}_\perp,{\vec k}_\perp)\\
L_{\cal U}&=& \int dx d^2{\vec b}_\perp d^2{\vec k}_\perp \left({\vec b}_\perp \times {\vec k}_\perp \right)^z
W^{\cal U}(x,{\vec b}_\perp,{\vec k}_\perp).
\nonumber
\end{eqnarray}
No issues with the Heisenberg uncertainty principle arise here since only perpendicular combinations of position ${\vec b}_\perp$ and momentum ${\vec k}_\perp$ are
needed simultaneously in order to evaluate the integral for
$L_{\cal U}$.

A straight line connecting $0$ and $\xi$ for the Wilson line in ${\cal U}_{0\xi}$ is often the most natural choice, resulting in
\begin{equation}
\langle {\vec k}^q_\perp\rangle_{straight}\equiv 
 \int dx d^2{\vec b}_\perp d^2{\vec k}_\perp {\vec k}_\perp 
W^{straight}(x,{\vec b}_\perp,{\vec k}_\perp)=
\frac{ \int d^3{\vec r}\langle PS | 
q^\dagger({\vec r})  \frac{1}{i}{\vec D} q({\vec r})|PS\rangle}
{\langle PS |PS\rangle},
\label{eq:kJi}
\end{equation}
which vanishes by time-reversal invariance \cite{collins1}.

However, depending on the context, other choices for the path in the Wilson link ${\cal U}$ should be made. Indeed, in the context
of TMDs probed in SIDIS the path should be taken to be a straight line to $x^-=\infty$
along (or, for regularization purposes, very close to) the light-cone. This particular choice ensures proper inclusion of the
Final State Interactions (FSI) experienced by the struck quark as it leaves the nucleon
along a nearly light-like trajectory in the Bjorken limit. However, a Wilson line to
$\xi^-=\infty$, for fixed ${\vec \xi}_\perp$ is not yet sufficient to render Wigner distributions
manifestly gauge invariant, but a link at $\xi^-=\infty$ must be included to ensure manifest
gauge invariance. While the latter may be unimportant in some gauges, it is crucial in
light-cone gauge for the description of TMDs relevant for SIDIS \cite{jifengTMD}. 

\begin{figure}
\unitlength1.cm
\begin{picture}(10,4)(5,22)
\includegraphics{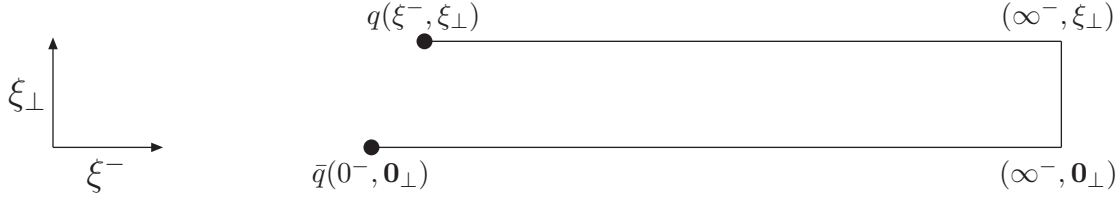}
\end{picture}
\caption{Illustration of the path for the Wilson line gauge link used to define the Wigner distribution $W^{+LC}$ 
(\ref{eq:wigner}).
}
\label{fig:staple}
\end{figure}

Let ${\cal U}^{+LC}_{0\xi}$ be the Wilson path ordered exponential obtained by first taking
a Wilson line from $(0^-,{\vec 0}_\perp)$ to $(\infty,{\vec 0}_\perp)$, 
then to $(\infty,{\vec \xi}_\perp)$, and then to $(\xi^-,{\vec \xi}_\perp)$, with each segment being a straight line (Fig. \ref{fig:staple}) \cite{hatta}. 
The shape of the segment at $\infty$ is irrelevant as the gauge field is pure gauge there, but it is still necessary to include a connection at $\infty$ and for
simplicity we pick a straight line. Likewise, with a similar 'staple' to $-\infty$ we define the Wilson path ordered exponential ${\cal U}^{-LC}_{0\xi}$, and using those light-like 
gauge links we define
\begin{equation}
W^{\pm LC}(k^+=xP^+,{\vec b}_\perp, {\vec k}_\perp)
\equiv \int \frac{d^2{\vec q}_\perp}{(2\pi)^2}\int \frac{d^2\xi_\perp d\xi^-}{(2\pi)^3}
e^{-i{\vec q}_\perp \cdot {\vec b}_\perp}
e^{i(xP^+\xi^-\!-{\vec k}_\perp\cdot{\vec \xi}_\perp)}
\langle P^\prime S^\prime |
\bar{q}(0)\Gamma {\cal U}^{\pm LC}_{0\xi}q(\xi)|PS\rangle .
\label{eq:wignerpm}
\end{equation}
This definition for $W^{+LC}$ the same as that in \cite{hatta} and similar to that of $W_{LC}$ in Ref. \cite{jifeng}, except that
the link segment at $x^-=\infty$ was not included in the definition of $W_{LC}$
\cite{jifeng}. The Wilson like gauge link used  to guarantee manifest gauge invariance is defined using a light-like 'staple, i.e. it is constructed using
three straight line gauge links\footnote{Some of the subtleties in regularizing/renormalizing such objects are addressed in Ref. \cite{Collins:what} and references therein.}
${\cal U}^{+ LC}_{0\xi} = W_{0^-0_\perp,\infty0_\perp}
W_{\infty0_\perp,\infty\xi_\perp}W_{\infty\xi_\perp,\xi^-\xi_\perp}$
and similarly for ${\cal U}^{- LC}_{0\xi}$.

If boundary conditions are chosen such that ${\vec A}_\perp(+\infty,{\vec r}_\perp)=0$, but ${\vec A}_\perp(-\infty,{\vec r}_\perp)\neq 0$ then
$W_{LC}$ from  Ref. \cite{jifeng} becomes equal to $W^{+LC}$. It turns out that the piece at $x^-=\infty$
is crucial for TMDs, but does not contribute to the OAM \cite{hatta}.

In light-cone gauge $A^+=0$ the Wilson lines to $x^-=\pm\infty$ become trivial and only the
piece at $x^-=\infty$ remains.
Although the gauge field at light-cone infinity ${\vec A}_\perp(\pm \infty,{\vec r}_\perp)$ cannot be neglected or set equal to zero in light-cone gauge,
it can be chosen to satisfy anti-symmetric boundary conditions
\begin{equation}
{\vec \alpha}_\perp({\vec r}_\perp)\equiv
{\vec A}_\perp(+\infty,{\vec r}_\perp) =-{\vec A}_\perp(-\infty,{\vec r}_\perp).
\label{eq:abc}
\end{equation}
This choice maintains manifest PT (sometimes called 'light-cone parity') invariance.

Using these Wigner distributions, one can now proceed to introduce the average transverse momentum as
\begin{eqnarray}
\langle {\vec k}^q_\perp \rangle_{\pm} &\equiv& \int dx d^2{\vec b}_\perp d^2{\vec k}_\perp {\vec k}_\perp 
W^{\pm LC}(x,{\vec b}_\perp,{\vec k}_\perp) \label{eq:kpm}\\
&=& \frac{ \int d^3{\vec r}\langle PS | 
\bar{q}({\vec r}) \gamma^+\left( \frac{1}{i}{\vec \partial}\mp g {\vec \alpha}_\perp ({\vec r}_\perp)\right)q({\vec r})|PS\rangle}
{\langle PS |PS\rangle}. \nonumber
\end{eqnarray}
Eq. (\ref{eq:kpm}) differs from (\ref{eq:kJi}) by the matrix element of (in $A^+=0$ gauge)
\begin{equation}
\bar{q}({\vec r}) \gamma^+\left[g{ A}^i_\perp ({\vec r}_\perp)-
g{\alpha}^i_\perp ({\vec r}) \right]q({\vec r})=- \bar{q}({\vec r}) \gamma^+\int_{r^-}^\infty dz^-
g\partial_- {A}^i_\perp (z^-,{\vec r}_\perp)q({\vec r})
= -\bar{q}({\vec r}) \gamma^+\int_{r^-}^\infty dz^- gG^{+i}(z^-,{\vec r}_\perp)q({\vec r}),
\label{eq:kp}
\end{equation}
where $G^{+\perp}=\partial_-A^\perp$ is the gluon field strength tensor in $A^+=0$ gauge. We note that for example
\begin{equation}
-\sqrt{2}gG^{+y}\equiv -gG^{0y}-gG^{zy} = g\left(E^y-B^x
\right)
=g\left({\vec E}+{\vec v}\times {\vec B}\right)^y
\end{equation}
yields the $\hat{y}$ component of the color Lorentz force acting on a particle that moves with the velocity of light in the $-\hat{z}$ direction (${\vec v}=(0,0,-1)$) --- which is the direction of the 
momentum transfer in DIS. Furthermore, the
integration of the matrix element of (\ref{eq:kp})
along the light-like trajectory of the ejected quark
yields the average change in momentum:
$\langle {\vec k}^q_\perp\rangle_{straight}=0$
while $\langle {\vec k}^\perp_q\rangle_{+LC}$ is
the $\perp$ momentum relevant for SIDIS experiments.
These observation motivate the semi-classical
interpretation of the matrix element of (\ref{eq:kp})
as  the average transverse momentum of the ejected quark as arising from the average color-Lorentz force from the
spectators as it leaves the target \cite{QS,mb:force}.

In a general gauge, there is an additional contribution from the
transverse derivative acting on the gauge links to/from $x^-=\infty$ and for example in an Abelian theory ${\vec \alpha}_\perp({\vec r}_\perp)$ in  (\ref{eq:kpm})
gets replaced by
\begin{equation}
{\alpha}^i_\perp	({\vec r}_\perp)\rightarrow 
{\alpha}^i_\perp	({\vec r}_\perp) - \int_{r^-}^\infty dz^-
{\partial}^i A^+ (z^-,{\vec r}_\perp) = 
{ A}^i_\perp
(r^-,{\vec r}_\perp)-
\int_{r^-}^\infty dz^- G^{+i}(z^-,{\vec r}_\perp),
\label{generalgauge}
\end{equation}
where $G^{+i}(z^-,{\vec r}_\perp) =
\partial_-A^i - \partial^i A^+$.
In the nonabelian case an additional
commutator as well additional gauge links connecting the quark and the gluon operators arise (see Section \ref{sec:gaugeinvariance}).
Eq. (\ref{generalgauge}) illustrates that the interpretation of the difference between the transverse momentum using
light-cone staples and that using straight-line gauge links
as the average color Lorentz force is gauge invariant.

The same Wigner distributions that we used to define average
transverse momentum can also be used to define OAM, yielding
\cite{jifeng}
\begin{equation}
L^q_{straight}\equiv 
 \int dx d^2{\vec b}_\perp d^2{\vec k}_\perp \left({\vec b}_\perp \times {\vec k}_\perp \right)^z
W^{straight}(x,{\vec b}_\perp,{\vec k}_\perp)=
\frac{ \int d^3{\vec r}\langle PS | 
q^\dagger({\vec r}) \left( {\vec r}\times \frac{1}{i}{\vec D}\right)q({\vec r})^z|PS\rangle}
{\langle PS |PS\rangle}=L^q_{Ji},
\label{eq:LJi}
\end{equation}
This is identical to the angular momentum that appears in the Ji-decomposition of the angular momentum for a nucleon (\ref{eq:JJi}).

Likewise, Wigner distributions employing light-like staples yield (in $A^+=0$ gauge)
\begin{eqnarray}
{\cal L}_{\pm}^q &\equiv& \int dx d^2{\vec b}_\perp d^2{\vec k}_\perp \left({\vec b}_\perp \times {\vec k}_\perp \right)^z
W^{\pm LC}(x,{\vec b}_\perp,{\vec k}_\perp) \label{eq:Lpm}\\
&=& \frac{ \int d^3{\vec r}\langle PS | 
\bar{q}({\vec r}) \gamma^+\left[{\vec r}\times\left( \frac{1}{i}{\vec \partial}\mp g {\vec \alpha}_\perp ({\vec r}_\perp\right)\right]^zq({\vec r})|PS\rangle}
{\langle PS |PS\rangle}, \nonumber
\end{eqnarray}
and similar for the glue. Eq. (\ref{eq:Lpm}) differs from 
\begin{equation}
{\cal L}^q = \frac{ \int d^3{\vec r}\langle PS | 
\bar{q}({\vec r}) \gamma^+\left({\vec r}\times \frac{1}{i}{\vec \partial}\right)^zq({\vec r})|PS\rangle}
{\langle PS |PS\rangle}
\end{equation}
(denoted $\tilde{L}^q$ in Ref. \cite{jifeng}) by the contribution from the gauge field $\mp {\vec \alpha}_\perp$ at $\pm \infty$. ${\cal L}^q$ is also identical to the quark OAM appearing in the Jaffe-Manohar
decomposition of the nucleon spin (\ref{eq:JJM}) as
we will discuss below.

\section{Connections Between Different Definitions for OAM}
First of all from  PT invariance one finds that ${\cal L}_+^q={\cal L}_-^q$ \cite{hatta}.
As a corollary,
since the piece at $\pm\infty$ cancels in the average
both must thus be identical to the OAM appearing in the Jaffe-Manohar decomposition
\begin{equation}
{\cal L}^q = \frac{1}{2}\left({\cal L}^q_{+}+{\cal L}^q_{-}\right)={\cal L}_+^q=
{\cal L}_-^q.
\end{equation}
Therefore, even though the gauge link at $x^-=\pm \infty$ is essential for the
description of TMDs \cite{jifengTMD}, it does not contribute to the OAM provided
anti-periodic boundary conditions (\ref{eq:abc}) in light-cone gauge are implied \cite{lorce2}.

To establish the connection with the orbital angular momentum entering the Ji-decomposition,
we consider (for simplicity in light-cone gauge)
\begin{eqnarray}
{\cal L}^q - L^q={\cal L}^q_+ - L^q &=&  \frac{ \int d^3{\vec r}\langle PS | 
\bar{q}({\vec r}) \gamma^+\left[{\vec r}_\perp \times\left( g {\vec A}_\perp ({\vec r})
-g{\vec \alpha}_\perp ({\vec r}_\perp)\right)\right]^zq({\vec r})|PS\rangle}
{\langle PS |PS\rangle}.
\end{eqnarray}
As discussed in Ref. \cite{BC}, we replaced $\gamma^0\rightarrow \gamma^+$ for a nucleon at rest in the definition for $L^q$.

Using (\ref{eq:kp},\ref{generalgauge}) and the semi-classical interpretation of $-gG^{+i}(r^-,{\vec r}_\perp)$ as the transverse Force acting on the active quark along its trajectory
we thus conclude that
\begin{equation}
T^z(r^-,{\vec r}_\perp)\equiv -g\left(x G^{+y}(r^-,{\vec r}_\perp)-yG^{+x}(r^-,{\vec r}_\perp)\right)
\end{equation}
represents the $\hat{z}$ component of the torque that acts on a particle moving with
(nearly) the velocity of light in the $-\hat{z}$ direction -- the direction in which the ejected
quark moves. Thus the difference between the (forward) light-cone definition $L^{+LC}=L_{JM}$ and the
local definition $L_{straight}=L_{Ji}$ of the orbital angular momentum is the
change in orbital angular momentum as the quark moves through the color field
created by the spectators
\begin{equation}
{\cal L}^q - L^q = \frac{ \int d^3{\vec r}\langle PS | 
\bar{q}({\vec r}) \gamma^+\int_{r^-}^{\infty}dz^- T^z(z^-,{\vec r}_\perp)
q({\vec r})|PS\rangle} {\langle PS |PS\rangle}.
\label{eq:torque}
\end{equation}
Therefore, while $L^q$ represents the local and manifestly gauge invariant OAM of the
quark {\it before} it has been struck by the $\gamma^*$, ${\cal L}^q$ represents 
the gauge invariant OAM {\it after} it has left the nucleon and moved to $r^-=\infty$.
This physical interpretation of the difference between the TMD based (i.e. Jaffe-Manohar) definition of quark OAM with a light-cone staple and the local definition represents the main  result of this paper.

\begin{figure}
\unitlength1.cm
\begin{picture}(10,8)(-1,12.7)
\includegraphics{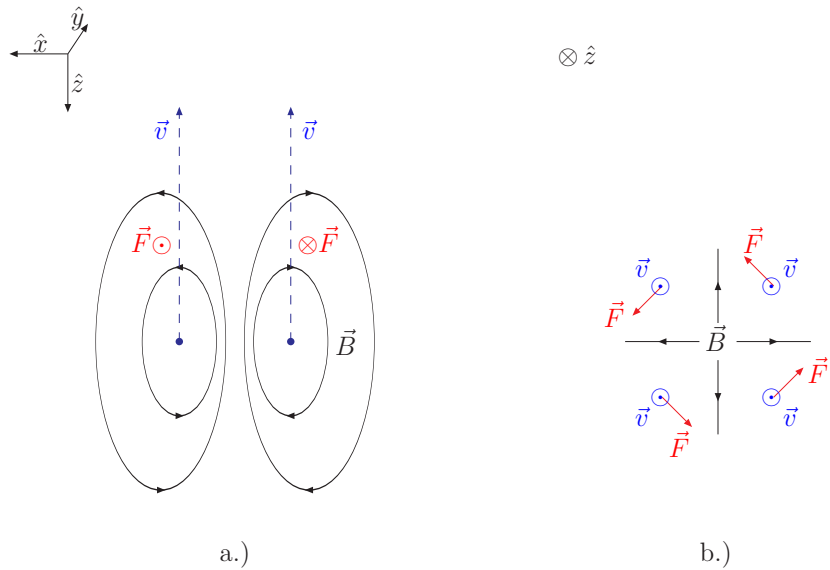}
\end{picture}
\caption{Illustration of the torque acting on the struck quark in the $-\hat{z}$ direction
through a color-magnetic dipole field caused by the spectators. a.) side view; b.) top view.
In this example the $\hat{z}$ component of the torque is negative as the quark leaves the
nucleon.
}
\label{fig:dipole}
\end{figure}


It is easy to see that a torque as appearing in (\ref{eq:torque}) may exist by considering the example of a quark moving through a (color-) magnetic dipole field caused by the spectators. Because of the overall color-neutrality, this is similar to a positively charged particle moving through the magnetic field caused by negative
spectators in QED. For spectator spins/OAMs that are oriented in the $+\hat{z}$ axis one would thus expect a dipole field as shown in Fig. \ref{fig:dipole}.
All quarks ejected in the $-\hat{z}$ direction pass through the region of outward pointing radial
magnetic field component, but only those originating in the bottom portion also move through
regions of inward pointing radial component, i.e. for quarks ejected in the $-\hat{z}$ direction
the regions of outward pointing radial component dominate.
One would thus expect more torque in the
$-\hat{z}$ direction than in $+\hat{z}$ direction. This example not only illustrates that the
net change in OAM as the quark leaves the nucleon is nonzero, but also suggests what the sign of
${\cal L}^q - L^q$ might be: for $d$ quarks the spins of the spectators
are positively correlated with the nucleon spin, corresponding to a situation similar to the
one depicted in Fig. \ref{fig:dipole}, and ${\cal L}^q - L^q$ should thus be negative.
For $u$ quarks the situation is less obvious since there should be a partial cancellation between
the $d$ quark spectator and the $u$ quark spectator. For an positron (electron) moving through its
own dipole field in QED  the magnetic dipole field is reversed. 
This illustrates why ${\cal L}^e - L^e$ is positive for an electron\cite{BC}.





\section{Gauge Invariance}
\label{sec:gaugeinvariance}
Although we started our investigation using manifestly gauge invariant Wigner functions (\ref{eq:wignerpm}), we picked light-cone gauge in order to arrive at simpler expressions for the OAM that allowed for a more direct
physical interpretation. However, for a complete discussion we provide manifestly
gauge invariant expressions for the quantities discussed in previous sections.

When one evaluates 
$L^{+LC}\equiv \int d^2k_\perp \int dx \int d^2\xi_\perp
W^{+LC}(x,{\vec k}_\perp,{\vec \xi}_\perp) \left( {\vec \xi}_\perp
\times {\vec k}_\perp\right)^z$, the factor ${\vec k}_\perp$ can be
translated into a $\perp$ derivative $-i\frac{\partial}{\partial \xi^i} \equiv -i \partial_i$ acting on the operator
$\bar{q}(0)\Gamma {\cal U}^{\pm LC}_{0\xi}q(\xi)$, whose
matrix element is subsequently
evaluated for ${\vec \xi}_\perp ={\vec 0}_\perp$.
The term where $-i\frac{\partial}{\partial \xi^i}$ acts on $q(\xi)$
yields the canonical OAM. More interesting is the term where
$-i\frac{\partial}{\partial r^i}$ acts on the staple-shaped gauge link
${\cal U}^{+ LC}_{0\xi} = W_{0^-0_\perp,\infty0_\perp}
W_{\infty0_\perp,\infty\xi_\perp}W_{\infty\xi_\perp,\xi^-\xi_\perp}$:
\begin{equation}
-i\left.\frac{\partial}{\partial \xi^i} {\cal U}^{+ LC}_{0\xi} 
\right|_{\xi =0} = 
W_{0^-0_\perp,\infty0_\perp}
A_i (\infty,0_\perp)
W_{\infty0_\perp,0^-0_\perp}
+
\int_{0^-}^{\infty} dz^-
W_{0^-0_\perp,z^-0_\perp}
\partial_iA^+(z^-0_\perp)W_{z^-0_\perp,0^-0_\perp},
\end{equation}
where the first term arises from the derivative acting on
the link at $\xi^-=\infty$, and the second when it acts on the
link from $\xi^-$ to $\infty$. Here we used that these path-ordered exponentials satisfy
\begin{equation}
-i\left.\frac{\partial}{\partial \xi^i}  
W_{\infty\xi_\perp,\xi^-\xi_\perp}\right|_{\xi=0} = 
\int_{0^-}^{\infty}dz^-
W_{\infty0_\perp,z^-0_\perp} \partial_iA^+(z^-0_\perp)
W_{z^-0_\perp,0^-0_\perp}
\end{equation}
and $W_{0^-0_\perp,\infty0_\perp}
W_{\infty0_\perp,z^-0_\perp}=W_{0^-0_\perp,z^-0_\perp}$.
Using integration by parts
\begin{eqnarray}
\int_{0^-}^{\infty}dz^-
W_{\infty0_\perp,z^-0_\perp} \partial_-A^i(z^-{\vec 0}_\perp)
W_{z^-0_\perp,0^-0_\perp}
&=& W_{\infty0_\perp,z^-0_\perp}
A^i(\infty,{\vec 0}_\perp)W_{z^-0_\perp,0^-0_\perp}
- A^i(0^-,{\vec 0}_\perp)\\
& &- \int_{0^-}^{\infty} dz^-W_{\infty0_\perp,z^-0_\perp} \left[A^-(z^-{\vec 0}_\perp),A^i(z^-{\vec 0}_\perp)\right]
W_{z^-0_\perp,0^-0_\perp}\nonumber
\end{eqnarray}
Thus
\begin{eqnarray}
-i\left.\frac{\partial}{\partial \xi^i} {\cal U}^{+ LC}_{0\xi} 
\right|_{\xi =0} +gA^i(0,{\vec 0}_\perp)=-
g\int_{0^-}^{\infty}dz^-W_{\infty0_\perp,z^-0_\perp} G^{+i} (z^-,{\vec 0}_\perp)W_{z^-0_\perp,0^-0_\perp},
\end{eqnarray}
where
\begin{equation}
G^{+i} =\partial_-A^i-\partial^iA^+ +ig\left[A^+,A^i\right].\label{eq:G+i}
\end{equation}
Inserting this result into our definition of Wigner functions one thus finds for the transverse momentum and the angular momentum, respectively
\be
\langle {\vec k}^q_\perp \rangle_{+}=
\frac{ \int d^3{\vec r}\langle PS | 
\bar{q}({\vec r}) \gamma^+\left(\frac{1}{i}
\partial_\perp -gA^i(0,{\vec 0}_\perp)-
\int_{0^-}^{\infty}dz^-W_{\infty0_\perp,z^-0_\perp} G^{+i} (z^-,{\vec 0}_\perp)W_{z^-0_\perp,0^-0_\perp}
\right)q({\vec r})|PS\rangle}
{\langle PS |PS\rangle}.
\ee
\be
\\ \nonumber
{\cal L}^q_{+LC}=
\frac{ \int d^3{\vec r}\langle PS | 
\bar{q}({\vec r})\gamma^+\left[x\!\left(\frac{1}{i}
\partial^y -gA^y(0,{\vec 0}_\perp)-
\int_{r^-}^{\infty}dz^-W_{r^- r_\perp,z^-r_\perp} gG^{+i} (z^-,{\vec r}_\perp)W_{z^-r_\perp,r^-r_\perp}
\right) - 'x\leftrightarrow y'\right]q({\vec r})|PS\rangle}
{\langle PS |PS\rangle}.
\ee
The difference \footnote{In the case of the transverse momentum, the local matrix element $\langle {\vec k}^q_\perp \rangle_{straight}=0$ 
vanishes due to time-reversal invariance \cite{collins1}, i.e.
$\langle {\vec k}^q_\perp \rangle_{+}-
\langle {\vec k}^q_\perp \rangle_{straight}=
\langle {\vec k}^q_\perp \rangle_{+}$ but we
discuss here the difference to facilitate the comparison with the OAM, where the local matrix
element ${\cal L}^q_{straight}$ does not vanish.}
\be
\langle {\vec k}^q_\perp \rangle_{+}-
\langle {\vec k}^q_\perp \rangle_{straight}=
-\frac{ \int d^3{\vec r}\langle PS | 
\bar{q}({\vec r}) \gamma^+
\int_{r^-}^{\infty}dz^-W_{r^- r_\perp,z^-r_\perp} gG^{+i} (z^-,{\vec r}_\perp)W_{z^-r_\perp,r^-r_\perp}
q({\vec r})|PS\rangle}
{\langle PS |PS\rangle}
\ee
is the well-known Qiu-Sterman matrix element
\cite{QS}
that
has the physical interpretation as the change in transverse momentum for the struck quark as it leaves the target after being struck by the virtual 
photon in a DIS experiment
. Semiclassically, that change in momentum is due to the color Lorentz
force as the quark leaves the target.

For the OAM one finds for the difference
\be
{\cal L}^q_{+LC}-{\cal L}^q_{straight}
=
-\frac{ \int d^3{\vec r}\langle PS | 
\bar{q}({\vec r}) \gamma^+\!\left[x
\int_{r^-}^{\infty}dz^-W_{r^- r_\perp,z^-r_\perp} gG^{+y} (z^-,{\vec r}_\perp)W_{z^-r_\perp,r^-r_\perp}- 'x\leftrightarrow y'\right]
q({\vec r})|PS\rangle}
{\langle PS |PS\rangle}
\ee

Since in light-cone gauge the light-like Wilson lines become trivial and (\ref{eq:G+i}) reduces to $
\partial_-A^i$ , matrix elements involving
the above the above correlation functions thus provide a manifestly gauge invariant extension of our key observation regarding the difference between the Jaffe-Manohar definition for quark orbital angular momentum and that of Ji.

The fact that the only difference between
$\langle {\vec k}^q_\perp \rangle_{+}$
and ${\cal L}^q_{+LC}$ is multiplication by a transverse position also illustrates that
any renormalization issues for ${\cal L}^q_{+LC}$
are similar to that of $\langle {\vec k}^q_\perp \rangle_{+}$ \cite{Collins:what}
. For example, as with $\langle {\vec k}^q_\perp \rangle_{+}$, issues with light-cone
singularities can be controlled by going slightly
off the light-cone in the case of ${\cal L}^q_{+LC}$
as well. Furthermore,
the same evolution equations that govern scale dependencies for $\langle {\vec k}^q_\perp \rangle_{+}$, should
also  describe that of ${\cal L}^q_{+LC}$
(multiplied by the appropriate transverse position factor).

\section{Summary}
The angular momenta appearing in the Jaffe-Manohar formalism are identical to Wigner function based definitions of OAM utilizing light-cone staples. We have used this result to
uderstand the difference between the Jaffe-Manohar definition of OAM and Ji's
local manifestly gauge invariant definition of OAM can be related to the torque that acts on a quark in longitudinally polarized DIS. In other words., while one definition (Ji) yields the net OAM quarks {\it before} absorbing the virtual photon, the (light-cone staple) Wigner distribution based definition (JM) yields the net OAM after the quark has escaped to infinity.
We thus now understand the physics through which  these two definitions are
related to one another. 

This is very similar to the situation in the context of TMDs where the difference between the average quark transverse momentum after it has left the target
(from Sivers function) and before it has left the target (where it is zero),
can be related to the difference of TMDs defined with a light-cone staple shaped
Wilson line gauge link versus one defined with a straight-line gauge link.

Unfortunately, no experiment has been identified to measure the OAM of quarks after they have been ejected in DIS. Nevertheless, we believe that the above interpretation
will help to develop a more complete picture of the nucleon spin. 

{\bf Acknowledgements:}
I would like to thank G. Schnell for stimulating 
discussions during the early stages of this work and Y. Hatta as well as C. Lorc\'e 
for very useful comments on the $1^{st}$ version of the paper.
This work was partially supported by the DOE under grant number 
DE-FG03-95ER40965. 
\bibliography{parton_oam5.bib}

\end{document}